\documentclass[pra,prl,twocolumn,superscriptaddress,floatfix,showpacs,10pt,letterpaper]{revtex4}

\usepackage{graphicx}
\usepackage{dcolumn}
\usepackage{bm}
\usepackage{subfigure}
\usepackage{amssymb}
\usepackage{verbatim}
\usepackage{amsmath}
\usepackage{amscd}
\usepackage{amssymb}

\begin{document}

\begin{titlepage}

\title{Gapped Two-body Hamiltonian whose Unique Ground State is \\ Universal for One-way Quantum Computation}

\author{Xie Chen}
\affiliation{Department of Physics, Massachusetts Institute of
Technology, Cambridge, Massachusetts 02139, USA}

\author{Bei Zeng}
\affiliation{Department of Physics, Massachusetts Institute of
Technology, Cambridge, Massachusetts 02139, USA}

\author{Zhengcheng Gu}
\affiliation{Department of Physics, Massachusetts Institute of
Technology, Cambridge, Massachusetts 02139, USA}

\author{Beni Yoshida}
\affiliation{Department of Physics, Massachusetts Institute of
Technology, Cambridge, Massachusetts 02139, USA}

\author{Isaac L. Chuang}
\affiliation{Department of Physics, Massachusetts Institute of
Technology, Cambridge, Massachusetts 02139, USA}

\begin{abstract}
Many-body entangled quantum states studied in condensed matter physics
can be primary resources for quantum information, allowing any quantum
computation to be realized using measurements alone, on the state.
Such a universal state would be remarkably valuable, if only it were
thermodynamically stable and experimentally accessible, by virtue of
being the unique ground state of a physically reasonable Hamiltonian
made of two-body, nearest neighbor interactions.  We introduce such a
state, composed of six-state particles on a hexagonal lattice, and
describe a general method for analyzing its properties based on its
projected entangled pair state representation.
\end{abstract}


\maketitle

\end{titlepage}

\def\>{\rangle}
\def\<{\langle}


Many-body entanglement is fundamental to the understanding of complex
condensed matter systems, as well as a primary resource for quantum
computation. A surprising result in quantum computation is that
certain entangled states can be employed to perform arbitrary quantum
information processing tasks, merely by systematically measuring
single sites in different bases, in a method known as ``one-way''
quantum computation\cite{1WQC}. If such universal resource states are
available, this approach potentially simplifies experimental
requirements by employing only measurements, and not multi-qubit gates
normally needed. Exotic physical properties may arise in these states
due to their large amount of entanglement; many methods have been
developed in condensed matter theory to study such systems, including
the matrix product state formalism \cite{MPS} or more generally the
projected entangled pair (PEPS) representation of states \cite{PEPS}.

The special entangled states which make arbitrary one-way quantum
computation possible unfortunately seem to be difficult to realize
naturally. Ideally, such universal resource states could be obtained
as the unique ground state of a naturally occurring Hamiltonian, one
with only nearest-neighbor two-body interactions. If this were the
case, especially if an energy gap existed between the ground and first
excited states, the one-way quantum computation could be robust
against quantum noise and decoherence of the entanglement.  However,
no such parent Hamiltonian exists for any of the presently known
resource states of one-way quantum computation.

Many efforts have been made to construct the desired many-body
entangled state such that it could be the ground state of a
naturally occurring Hamiltonian. The first and best known resource
state is the cluster state, a simple entangled state on a
two-dimensional square lattice; unfortunately, it cannot be the
exact ground state of any naturally occurring
Hamiltonian\cite{Graph}. Perturbative approaches providing a
Hamiltonian whose ground state approximates that desired have been
developed \cite{Graph,Sta_H,Barlett}. A nice scheme for constructing
universal resource states has been proposed and has yielded many
interesting examples\cite{Gross}. Based on this, a mixed approach
can be taken, using a 1D Hamiltonian to create chains, that are then
coupled by two-body unitary operations\cite{Gross,Brennen} to form a
2D resource state. Matrix product state\cite{MPS} techniques allow
any measurement of these 1D chains to be computed efficiently, on a
classical computer, however, implying that they alone are
insufficient for quantum computation. Two-dimensional many-body
entangled states are thus likely to be essential for arbitrary
quantum computations, but few techniques are presently known for
finding local 2D Hamiltonians with the desired ground states.
Properties of such states generally remain intrinsically hard to
determine \cite{Complex}.

Here, we present results from a new approach to studying the quantum
informational and physical properties of 2D many-body entangled states
using the PEPS representation.  On the one hand, this representation
naturally includes many-body entanglement in its state description\cite{PEPS} and
hence facilitates understanding of one-way quantum computation
schemes\cite{VBS,Gross}. On the other hand, methods have been developed to study the
physical properties of PEPS states as ground state of parent
Hamiltonians\cite{PEPS_H}. Combining these insights, we are able to construct the
first example of a system which is both the unique ground state of a
gapped two-body nearest-neighbor Hamiltonian and a universal resource
state for one-way quantum computation.

\begin{figure}[htb!]
\centering \includegraphics[width=3in]{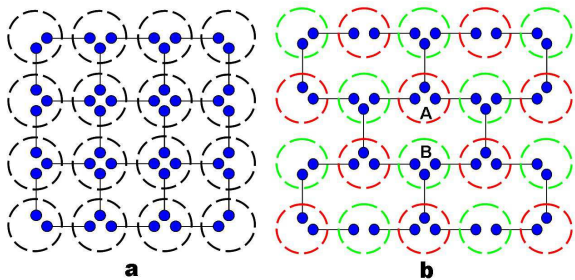}
\caption{Projected Entangled Pair State (PEPS) representation of 2D
(a) square $|\psi_{\rm PEPS}^{Sqr}\>$ and (b) hexagonal
$|\psi_{\rm PEPS}^{Hex}\>$ lattice states. Filled circles
connected by solid lines denote virtual singlet pairs
$|\varphi\>$. Dashed circles denote projection of virtual qubits
into physical states; red and green ones correspond to sublattices A
and B in hexagonal lattice respectively.}\label{fig:PEPS_l}
\end{figure}

\textbf{Building on PEPS} -- A good place to start in constructing the
desired state is with an example which illustrates the challenge,
based on the well-known cluster state. Consider the state $|\psi_{\rm
PEPS}^{Sqr}\>$ defined on a square lattice (Fig.~\ref{fig:PEPS_l}a)
where each pair of nearest-neighbor sites are connected by singlets
$|\varphi\> = |00\>+|01\>+|10\>-|11\>$ (suppressing normalization). On
sufficiently large lattices, starting with $|\psi_{\rm PEPS}^{Sqr}\>$,
any quantum circuit can be efficiently simulated by measuring all four
qubits at each site (on the boundary, two or three) in appropriate
time sequences and measurement bases\cite{VBS}.

$|\psi_{\rm PEPS}^{Sqr}\>$ is the unique ground state of a gapped
two-body Hamiltonian, as it is simply a tensor product of two-body
entangled states.  However, the multi-particle measurement required to
make this state universal\cite{VBS} is generally disallowed in one-way quantum
computation models.  Still, if the four qubits at each site were
treated as a single $16$ dimensional particle, the model could be
interpreted as giving the desired result, a universal resource state
for one-way quantum computation with a gapped two-body parent
Hamiltonian.  And while use of $16$ dimensional particles is
experimentally unrealistic, the idea of using a description in terms
of singlet pairs does provide a good starting point for constructing
simpler states.

Specifically, consider the set of states given by projecting
lattices of singlets into smaller subspaces.  For example, the
projector $P_{Cluster}^{Sqr}=|\tilde{0}\>\<0000| +
|\tilde{1}\>\<1111|$, applied to all sites of the square lattice
state gives the cluster state on a square
lattice \cite{VBS}, $|\Psi_{Cluster}\> \propto P_{Cluster}^{Sqr}
|\psi_{\rm PEPS}^{Sqr}\>$, where $|\tilde{0}\>$ and $|\tilde{1}\>$
are the physical qubits in the cluster state model.  In this ``PEPS
representation'' picture, the physical PEPS state is defined by two
elements, a lattice of ``virtual'' singlets (connecting neighboring
sites in the lattice), and a set of projectors which act on lattice
sites. Not all PEPS states are universal for quantum computation;
only a few, such as $|\Psi_{Cluster}\>$, are known to be universal.

Compared with $|\psi_{\rm PEPS}^{Sqr}\>$ with 16-dimensional
particles, $|\Psi_{Cluster}\>$ employs only qubits at each
site, and hence is more experimentally accessible.  Unfortunately it
cannot occur as the exact ground state of nearest-neighbor
interactions \cite{Graph}, and its parent Hamiltonian involves at
least five-body interactions.  Moreover, it is known that PEPS states
generally disallow low dimensionality particles simultaneously with
short interaction ranges \cite{PEPS_H}.  Nevertheless, this line of
thought, using PEPS states, can indeed lead to a universal resource
state which is the unique ground state of a gapped nearest-neighbor
Hamiltonian, while also being composed of particles of relatively low
dimension, as we now show.

\textbf{The tri-Cluster State} -- The structure of the
lattice of singlets, and the choice of projectors, in the construction
of PEPS states, provide powerful degrees of freedom for exploring
interesting new states.  Two specific insights from the above examples
illustrate this freedom:

{\noindent \bf 1.} Instead of on a square lattice, a cluster state
defined on a {\em hexagonal} lattice $|\Psi^{Hex}_{Cluster}\>$ is also
universal \cite{MVDN}. On a hexagonal lattice of singlet pairs
(Fig.~\ref{fig:PEPS_l}b), the projector defining this cluster state is
$P_{Cluster}^{Hex} = |\tilde{0}\>\<000| + |\tilde{1}\>\<111|$, giving
$|\Psi^{Hex}_{Cluster}\> \propto P_{Cluster}^{Hex} |\psi_{\rm
PEPS}^{Hex}\>$, where the labels denote left-right-up and
left-right-down virtual qubits on sites in sublattices A and B,
respectively. 

{\noindent \bf 2.} An alternative projector can be chosen:
$P'=|\tilde{0}\>\<100|+|\tilde{1}\>\<011|$ or
$P''=|\tilde{0}\>\<010|+|\tilde{1}\>\<101|$; these result in PEPS
states different from $|\Psi^{Hex}_{Cluster}\>$, but only by local Pauli
operations.  Hence, a modified local measurement scheme still
exists, allowing these states to also be universal.

We now introduce a new state, the {\em tri-Cluster} state
$|\Psi_{triC}\>$, which is motivated by these two insights, and
has properties we desire.  This is defined in the PEPS representation
on a two-dimensional hexagonal lattice (Fig.~\ref{fig:PEPS_l}b), with
projectors
\begin{eqnarray}
P_{triC}& = &  |\tilde{0}\>\<000| + |\tilde{1}\>\<111| \nonumber\\
        & + &  |\tilde{2}\>\<100| + |\tilde{3}\>\<011| \nonumber\\
        & + &  |\tilde{4}\>\<010| + |\tilde{5}\>\<101|
\label{P6}
\,,
\end{eqnarray}
using the same labeling scheme as above, such that $|\Psi_{triC}\>
\propto P_{triC} |\psi_{\rm PEPS}^{Hex}\>$.  Hence, at each lattice
site there lives a 6-dimensional particle.

Intuitively, $|\Psi_{triC}\>$ is universal because it is closely
related to the standard cluster state.  Specifically, $|\Psi_{triC}\>$
projected into the subspace spanned by $\{|\tilde{0}\>$,
$|\tilde{1}\>\}$ is the same as $|\Psi_{Cluster}\>$, as are also the
states given by $|\Psi_{triC}\>$ projected into
$\{|\tilde{2}\>$,$|\tilde{3}\>\}$ and
$\{|\tilde{4}\>$,$|\tilde{5}\>\}$, up to local Pauli errors.  Thus,
$|\Psi_{triC}\>$ is like a ``superposition'' of three cluster
states. Computational qubits are encoded in the virtual qubits and operated upon by measuring the physical particles. Although the three subspaces of $|\Psi_{triC}\>$ cannot be
decoupled physically, they may be employed independently in processing encoded
qubits with a suitable choice of measurement basis, as detailed later.

The most interesting nontrivial feature of $|\Psi_{triC}\>$ is that it
is the unique ground state of a gapped two-body Hamiltonian, and we
begin with that.

\textbf{Uniqueness \& Gap} --
The fact that $|\Psi_{triC}\>$ occurs as the unique ground state of a
gapped two-body Hamiltonian is very surprising, as on the one hand the
ground states of two-body Hamiltonians are rarely exactly known and on
the other hand simply constructed states do not usually have simple
parent Hamiltonians.  Even the one-dimensional cluster state requires
3-body interactions in its parent Hamiltonian. Below, we give a
two-body nearest-neighbor Hamiltonian $H_{triC}$ which has
$|\Psi_{triC}\>$ as its ground state.  Furthermore, we prove that
$|\Psi_{triC}\>$ is the only ground state of $H_{triC}$ and the
Hamiltonian has a constant gap independent of system size.

\begin{figure}[htb!]
\centering \includegraphics[width=1.5in]{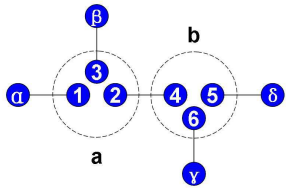}
\caption{One representative site with particles $a$ and $b$, and
neighboring boundary, in the hexagonal lattice of $|\psi_{\rm
PEPS}^{Hex}\>$.  Filled circles connected by solid lines represent
virtual singlets $|\varphi\>$ and dashed circles indicate sites
projected to obtain the physical state.}
\label{fig:2body}
\end{figure}

The central step in constructing $H_{triC}$ and studying its
properties is to find the support space $S_{ab}$ of the reduced
density matrix of any two nearest-neighbor particles $a$ and $b$ in
the state ($a$, $b$ are in two sublattices $A$, $B$ respectively).
This is accomplished by first finding the corresponding support
space $S_{ab}^{\rm PEPS}$ of the six virtual qubits on site $a$ and
$b$, in the PEPS picture, and then computing $S_{ab} \propto
P_{triC}S_{ab}^{\rm PEPS}$. For example, when $a$ is to the left of $b$
(Fig.~\ref{fig:2body}),
virtual qubits 1 to 6 on those sites are only connected to virtual
qubits $\alpha$, $\beta$, $\gamma$, $\delta$ elsewhere. By tracing out
$\alpha$ to $\delta$ from the 5 singlet pairs, we find $S_{ab}^{\rm
PEPS}$ for virtual qubits 1 to 6 to be spanned by
$|\pm\>_1|\pm\>_3|\varphi\>_{24}|\pm\>_5|\pm\>_6$, where $|\pm\>=(|0\>\pm |1\>)/\sqrt{2}$ and $|\varphi\>$ is the singlet state. This
16-dimensional space is then projected to give $S_{ab}$ for the
depicted lattice site.  $S_{ab}$ is different for the three bond
directions in a hexagonal lattice, i.e. $a$ to the left of, to
the right of, and below $b$.

Providing a two-body Hamiltonian with $|\Psi_{triC}\>$ being a
ground state is straightforward.  The Hilbert space of two
neighboring sites $a$, $b$ is 36-dimensional, larger than the
dimension of $S_{ab}$. Therefore we may choose any non-negative
Hermitian operator $h_{ab}$ on the two sites that has $S_{ab}$ as
its null space, such that $h_{ab} |\Psi_{triC}\> = 0$ for every
$h_{ab}$. Thus, $|\Psi_{triC}\>$ is {\em a} ground state of the
two-body Hamiltonian $H_{triC} = \sum_{ab} h_{ab}$, where the
summation is over all nearest-neighbor pairs. However, the key is to
construct $H_{triC}$ such that $|\Psi_{triC}\>$ is the {\em unique}
ground state, and it turns out the above procedure does work.

Specifically, let $h_{ab}$ be the projection operator $h^p_{ab}$ which
projects onto the $36-16 = 20$ dimensional subspace orthogonal to
$S_{ab}$, giving the total Hamiltonian
\begin{equation}
  H_{triC} = \sum_{a\in A} \left(h^p_{ab} + h^p_{ba} +
         h^p_{\stackrel{b}{a}}\right)
\label{H}
\,.
\end{equation}
The summation is over sites $a$ in sublattice $A$ and the three terms
$h^p_{ab}$, $h^p_{ba}$, $h^p_{\stackrel{b}{a}}$ correspond,
respectively, to three bond directions where $a$ is to the left, to
the right and below $b$. The Hamiltonian is hence
invariant under translation along sublattice $A$.  An explicit
expression for $H_{triC}$ in terms of spin operators can be given (see
appendix).

The specific $H_{triC}$ we have presented has $|\Psi_{triC}\>$ as its
unique ground state.  This is shown by verifying the condition
\cite{PEPS_H} that for any region $R$ of spins in $|\Psi_{triC}\>$,
the support space $S_R$ of the reduced density matrix on $R$ satisfies
\begin{equation}
    S_R = \bigcap_{\< ab \>} S_{ab}\otimes I_{R\setminus ab}
\,,
\label{Unic}
\end{equation}
where the intersection is taken over all neighboring pairs $ab$ and
$I_{R\setminus ab}$ is the full Hilbert space of all spins in region
$R$ except $a$ and $b$. For every possible configuration containing three or four
connected sites in $|\Psi_{triC}\>$ the condition is confirmed by direct
calculation. To check the condition for larger regions, it is useful to notice that any region in $|\Psi_{triC}\>$ containing more than one site is {\em injective}
\cite{PEPS_H}. By Lemma 2 of \cite{PEPS_H}, 1. if regions $R_1$ and $R_3$ are
not connected and $R_2$ and $R_3$ are injective, then $S_{R_1 \cup R_2
\cup R_3} = (S_{R_1 \cup R_2}\otimes I_{R_3})\cap (S_{R_2 \cup
R_3}\otimes I_{R_1})$ 2. if regions $R_1$, $R_2$, $R_3$ are all
injective, then $S_{R_1 \cup R_2 \cup R_3} = (S_{R_1 \cup R_2}\otimes
I_{R_3})\cap (S_{R_2 \cup R_3}\otimes I_{R_1}) \cap (S_{R_1 \cup
R_3}\otimes I_{R_2})$.  Hence for a region $R$ containing more than 4
sites in $|\Psi_{triC}\>$, $S_R$ is the intersection of all four-body
support spaces in $R$.  By induction, it follows that condition
Eq.(\ref{Unic}) is satisfied on $|\Psi_{triC}\>$ for any
$R$. Therefore, $|\Psi_{triC}\>$ is the unique ground state of $H_{triC}$.

$H_{triC}$ is also gapped; an energy gap $\eta$ above the ground state
exists, which is constant as the system size goes to infinity.  The
existence of this gap guarantees protection of $|\Psi_{triC}\>$
against thermal noise, independent of system size.  $\eta$ can be
bounded.  First, we show that $\eta$ is greater than $\lambda$, the
gap of another Hamiltonian $K$ which also has $|\Psi_{triC}\>$ as its
unique ground state, but has four-body terms instead of only two-body
terms.  We then bound $\lambda$ above a positive constant value.

\begin{figure}[htb!]
\centering
\includegraphics[width=2.0in]{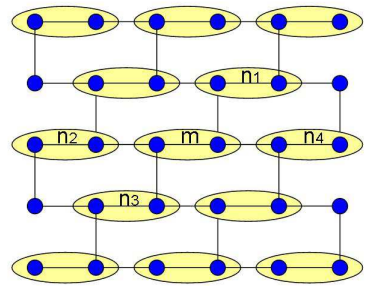}
\caption{Regrouping of lattice sites in tri-Cluster State into
disjoint blocks, each containing two sites.} \label{fig:HX}
\end{figure}

Consider the Hamiltonian $K$ for a re-labeled version of
$|\Psi_{triC}\>$, in which particles are regrouped into disjoint
blocks each containing two nearest-neighbors (Fig.~\ref{fig:HX}).
Let $K = \sum_{mn} k_{mn}$, where $m,n$ denote two connected
blocks, each containing two particles $m^{[l]},m^{[r]}$ and
$n^{[l]},n^{[r]}$ respectively, and $k_{mn}$ is projection onto the
orthogonal space of the four-body reduced density matrix on
$m^{[l]},m^{[r]},n^{[l]},n^{[r]}$ (assuming $m^{[r]}$ and $n^{[l]}$
are connected). Then $H_{triC} = \sum_{ab} h^p_{ab} \ge \frac{1}{4}
\sum_{mn} \left(h^p_{m^{[l]}m^{[r]}} + h^p_{m^{[r]} n^{[l]}} +
h^p_{n^{[l]}n^{[r]}}\right) \ge \frac{1}{4} \sum_{mn} \mu k_{mn} =
\frac{1}{4}\mu K$. $\left(h^p_{m^{[l]}m^{[r]}} + h^p_{m^{[r]} n^{[l]}}
+ h^p_{n^{[l]}n^{[r]}}\right)$ and $k_{mn}$ are both non-negative
operators with the same null space, so the last inequality holds for
some positive number $\mu$.  Assume that the gaps of the projectors
$h^p_{ab}$ and $k_{mn}$ are both $1$.  Direct calculation gives
$\mu=\frac{1}{2}$.  As discussed in the uniqueness proof, $K$ also
has $|\Psi_{triC}\>$ as its unique ground state.  Using this, we find
$\eta \ge \frac{1}{4}\mu \lambda = \frac{1}{8} \lambda$.
The gap $\lambda$ can be bounded by showing that $K^2 \ge c K$ for
some positive constant $c$. $K^2 = (\sum_{mn} k_{mn})^2 = K +
\sum_{mn,m'n'} (k_{mn} k_{m'n'} + k_{m'n'} k_{mn}) \ge K +
\sum_{n_imn_j} k_{mn_i} k_{mn_j} + k_{mn_j} k_{mn_i}$, $n_i$ and $n_j$
are blocks connected to $m$. The last inequality holds because when
region $mn$ and region $m'n'$ do not intersect $k_{mn} k_{m'n'} +
k_{m'n'} k_{mn} \ge 0$. Direct calculation shows that
(Fig.~\ref{fig:HX}) $k_{mn_i} k_{mn_j} + k_{mn_j} k_{mn_i} \ge 0$ for
$(i,j) = (1,2),(1,3),(2,4) \ \text{or} \ (3,4)$ and $k_{mn_i} k_{mn_j}
+ k_{mn_j} k_{mn_i} \ge -\frac{1}{3}k_{mn_i} -\frac{1}{3}k_{mn_j}$ for
$(i,j) = (1,4) \ \text{or}\ (2,3)$.  Summing over all consecutive
$n_i$, $m$, and $n_j$ gives $\sum_{n_imn_j} k_{mn_i} k_{mn_j} +
k_{mn_j} k_{mn_i} \ge -\frac{2}{3}\sum_{mn} k_{mn}$.  Therefore $K^2
\ge \frac{1}{3}K$, giving $\lambda \ge \frac{1}{3}$. Finally, we find
a lower bound on the gap $\eta$ of $H_{triC}$ of $\eta \ge \frac{1}{8}
\lambda \ge \frac{1}{24}$.


\textbf{Universality} -- $|\Psi_{triC}\>$ is a universal resource
state, because of properties it inherits from the cluster state.
Similar to a cluster state, computational qubits are encoded in the virtual qubits, and the active computational state flows along the lattice
as measurements on the physical states are performed. In contrast, however, with
$|\Psi_{triC}\>$ extra Pauli errors occur, thus necessitating
additional analysis.  Below, we describe the different steps
necessary, focusing on initialization and readout, one-qubit gates,
and a two-qubit gate sufficient for universality.

{\em Initialization and readout}: Just as with the cluster state, with
$|\Psi_{triC}\>$, measurement in the six-state basis, $\{|\tilde{0}\>
\cdots |\tilde{5}\>\}$ accomplishes several tasks. First, such
measurement detaches unnecessary sites from their neighbors (up to a
known Pauli error). Next for state initialization, it gives a
post-measurement state with an encoded qubit projected into $|+\>$
(when the outcome is $\tilde{0}$, $\tilde{3}$ or $\tilde{4}$) and
$|-\>$ (for outcomes $\tilde{1}$, $\tilde{2}$ or $\tilde{5}$).  At the
end of computation, the encoded qubit can also be read out in this
way, giving $0$ (for $\tilde{0}$, $\tilde{2}$ or $\tilde{5}$), and $1$
(for $\tilde{1}$, $\tilde{3}$ or $\tilde{4}$).

{\em One-qubit gates}: Similar to gate implementations with the
cluster state, once a line in the lattice has been detached from the
rest, appropriately measuring a site in the line performs a single
qubit rotation, up to a known Pauli error.  Specifically, measuring in
the basis \{$|\tilde{0}\>\pm e^{i\theta}|\tilde{1}\>$,
$|\tilde{2}\>\pm e^{i\theta}|\tilde{3}\>$, $|\tilde{4}\>\pm
e^{-i\theta}|\tilde{5}\>$\} implements operation \{$HZ(\theta)$,
$XHZ(\theta)$,$ZHZ(\theta)$,$YHZ(\theta)$,$ZHZ(\theta)$,$YHZ(\theta)$\},
respectively, on the encoded qubit (using standard notation for qubit
gates, with $Z(\theta)$ denoting a rotation about $\hat{z}$ by angle
$\theta$), up to pre-existing Pauli frame errors from detaching the
line.

{\em Two-qubit controlled-$Z$ gate}: Measurement of two vertically
connected particles $a$ and $b$ implements the final ingredient needed
for universality, a controlled-$Z$ gate $CZ_{ab}$, just as with the
cluster state scheme, but with some additional Pauli frame errors.
Specifically, measuring in basis $\{\hat{0}\ldots \hat{5}\} =
\{|\tilde{0}\>\pm |\tilde{1}\>$, $|\tilde{2}\>\pm |\tilde{3}\>$,
$|\tilde{4}\>\pm |\tilde{5}\>\}$ implements the two-qubit operation
$\left(X_a^{u_a}Z_a^{v_a}H_a\right) \left(X_b^{u_b}Z_b^{v_b}H_b\right)
\otimes \left(X_a^{w_a}X_b^{w_b}CZ_{ab}X_a^{w_a}X_b^{w_b}\right)$ on
the two adjacent encoded qubits. For $x\in\{a,b\}$, $u_x=1$ for $x$
measurement outcomes $\hat{1}$, $\hat{3}$, or $\hat{5}$; $v_x=1$ for
outcomes $\hat{2}$, $\hat{3}$, $\hat{4}$, or $\hat{5}$; $w_x=1$ for
$\hat{4}$, $\hat{5}$; and $u_x,v_x,w_x$ are $0$ otherwise.  Much like
for the cluster state, when embedded in a larger circuit, more
complicated configurations arise in implementing a controlled-$Z$ gate
(see appendix), but the principles of propagating a Pauli frame remain
the same.

\textbf{Conclusion} -- $|\Psi_{triC}\>$ is a remarkable entangled
many-body state which is both universal for one-way quantum
computation and the unique ground state of a gapped Hamiltonian
$H_{triC}$, made of local two-body terms.  While imperfect, due to use
of six-state spins, it steps far closer to physical realizability than
previous models.  Moreover, the methods introduced here, based on the
PEPS representation, are very general.  These analysis methods lead
directly to a number of additional universal states, and deepen
connections between the study of complex condensed matter systems and
quantum information science.

\clearpage

\section*{Appendix}

\subsection*{A: Explicit Hamiltonian with spin operators } 

$|\Psi_{triC}\>$ is the ground state of a Hamiltonian $H_{triC}$,
which can be expressed explicitly as being a sum over two-body
nearest-neighbor interactions between six-state spins on a hexagonal
lattice.  Below, we give such an explicit expression with spin
operators, as $H^{\star}_{triC}$.  Let the spin operators for
particles $a$ and $b$ along directions $x$, $y$ and $z$ be $S_{a_x}$,
$S_{a_y}$, $S_{a_z}$ and $S_{b_x}$, $S_{b_y}$, $S_{b_z}$
respectively. $S_+ = S_x + iS_y$ and $S_-=S_x-iS_y$ are the raising
and lowering spin operators.  $H^{\star}_{triC}$ is translationally
invariant along sublattice $A$ and is a sum over three sets of local
terms at every site $a$ in $A$:
\begin{equation}
    H^{\star}_{triC} = \sum_{a} \left(h_{ab} + h_{ba} 
	+ h_{\stackrel{b}{a}}\right)
\,.
\label{H_S}
\end{equation}
$h_{ab}$, $h_{ba}$ and $h_{\stackrel{b}{a}}$ describe interactions
along the three bond directions at each site $a$, where $b$ is to the
right, to the left and above $a$ respectively. They can be expressed
explicitly as:
\begin{equation}
\begin{array}{rl}
h_{ab} & =  \nonumber\\
       & 2(2S_{a_z}-5)(2S_{a_z}-3)(2S_{a_z}-1)(2S_{a_z}+1)(4S_{a_z}+11) \nonumber \\
       & (2S_{b_z}+5)(2S_{b_z}+3)(2S_{b_z}-1)(2S_{b_z}+1)(4S_{b_z}-11) \nonumber \\
      -& 75\sqrt{2}S_{a_+}(2S_{a_z}-5)(2S_{a_z}+3)(2S_{a_z}-1)(2S_{a_z}+1) \ \\
       & (48S^4_{b_z}+64S^3_{b_z}-280S^2_{b_z}-272S_{b_z}+67) \nonumber \\
      +& 75\sqrt{2}(48S^4_{a_z}-64S^3_{a_z}-280S^2_{a_z}+272S_{a_z}+67) \nonumber \\
       & S_{b_+}(2S_{b_z}-5)(2S_{b_z}-3)(2S_{b_z}-1)(2S_{b_z}+3)\nonumber \\
      +& 4\sqrt{10}S^3_{a_+}(2S_{a_z}-1)(2S_{a_z}-3)\times \nonumber \\
       & (128S^5_{b_z}+560S^4_{b_z}-2840S^2_{b_z}-3848S_{b_z}+675) \nonumber \\
      +& 4\sqrt{10}(128S^5_{a_z}-560S^4_{a_z}+2840S^2_{a_z}-3848S_{a_z}-675)  \nonumber \\
       & S^3_{b_+}(2S_{b_z}-5)(2S_{b_z}-3) + h.c.
\end{array}
\,,
\end{equation}
where $h.c.$ denotes the Hermitian conjugate, as usual.  $h_{ba}$ can
be obtained by exchanging $a$, $b$ in the above.
$h_{\stackrel{b}{a}}$ is:
\begin{equation}
\begin{array}{rl}
h_{\stackrel{b}{a}} &= \nonumber\\
                    & -25(2S_{a_z}-5)(2S_{a_z}-3)(2S_{a_z}+3)(2S_{a_z}+5)\nonumber \\
                   +& 25S^3_{a_+}(2S_{a_z}-5)(2S_{a_z}-1) \nonumber \\
                    & (224S^5_{b_z}-16S^4_{b_z}-1968S^3_{b_z}+40S^2_{b_z}+3550S_{b_z}-9) \nonumber \\
                   -& 12S^5_{a_+} \nonumber \\
                    & (416S^5_{b_z}-80S^4_{b_z}-3600S^3_{b_z}+520S^2_{b_z}+5994S_{b_z}-125)\nonumber \\
                   +& h.c. + (a\Leftrightarrow b)
\,,
\end{array}
\end{equation}
where $(a\Leftrightarrow b)$ denotes an exchange of $a$ and $b$ in the
preceeding expression.  Because each positive-semidefinite local term
in $H^{\star}_{triC}$ has the same null space as $h^p_{ab}$,
$H^{\star}_{triC}$ is gapped and has $|\Psi_{triC}\>$ as its unique
ground state for the same reasons as for $H_{triC}$.

\subsection*{B: Two-qubit gate operation: example}

The hexagonal lattice structure of $|\Psi_{triC}\>$ can complicate
implementation of controlled-Z gates when embedded in larger
circuits.  For example, if computational qubits are encoded only in
every other line, then extra steps are needed to bring two qubits
together for the two-qubit gate.  Below, we illustrate how this can be
done, analogous to the complicated case presented in the original
cluster state paper\cite{1WQC}.

\begin{figure}[htb!]
\centering
\includegraphics[width=2in]{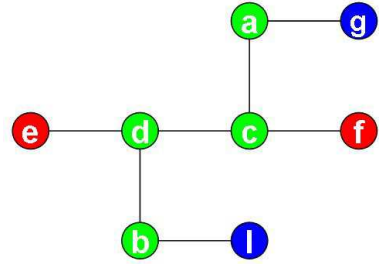}
\caption{Measurement pattern for simulating control-Z gate. $a$,$b$
are input sites and $g$, $l$ are output sites. Green sites are
measured in basis $\{|\tilde{0}\> \pm |\tilde{1}\>, |\tilde{2}\> \pm
|\tilde{3}\>, |\tilde{4}\> \pm |\tilde{5}\>\}$. Red sites $e$, $f$
are measured in computational basis
$\{|\tilde{0}\>,\dots,|\tilde{5}\>\}$ and their results may also
contribute error to the operation simulated.} \label{fig:CZ}
\end{figure}

Consider the configuration shown in Fig.~\ref{fig:CZ}, and suppose
information is made to propagate from left to right and the top and
bottom lines connect only through sites $c$ and $d$.  The initial
state $|\Phi_{ab}\>$ is input at sites $a$ and $b$.  Red sites $e$ and
$f$ are detached from the rest by measuring them in the six-state
basis, $\{|\tilde{0}\>,\dots,|\tilde{5}\>\}$, giving measurement
results $j_e$ and $j_f$.  The different measurement results obtained
contribute different Pauli errors to the two-qubit operation done
next.  Measuring the green sites $a$, $b$, $c$ and $d$ in basis
$\{\hat{0}...\hat{5}\} = \{|\tilde{0}\> \pm |\tilde{1}\>, |\tilde{2}\>
\pm |\tilde{3}\>, |\tilde{4}\> \pm |\tilde{5}\>\}$, gives results
$i_a$, $i_b$, $i_c$, $i_d$, and performs the two-qubit operation
operation $H_a H_b \otimes CZ_{ab}$ on $|\Phi_{ab}\>$ up to some Pauli
operation, leaving the final states output at sites $g$ and $l$.

Specifically, the operation performed, with all Pauli errors included,
is $H_{a} Z_{a}^{u_{c}+u_{e}+v_{d}} H_{b} Z_{b}^{u_{d}+u_{f}+v_{c}}
X_{a}^{u_{a}}  X_{b}^{u_{b}} CZ_{ab}
X_{a}^{v_{a}} Z_{a}^{w_{a}} X_{b}^{v_{b}} Z_{b}^{w_{b}}$,
where $v_{x}=1$ when $i_{x}=\hat{4}$ or $\hat{5}$;
$w_{x}=1$ when $i_{x}=\hat{1}$, $\hat{3}$, or $\hat{5}$;
$u_{x}=1$ when $i_{x}=\hat{4}$ or $\hat{5}$ (with $x$ being either $a$
or $b$); and
$u_{f}=1$ when $j_{f}=\tilde{1}$, $\tilde{2}$, or $\tilde{5}$;
$u_{e}=1$ when $j_{e}=\tilde{0}$, $\tilde{3}$, or $\tilde{4}$;
$u_{c}=1$ when $i_{c}=\hat{2}$, or $\hat{3}$;
$u_{d}=1$ when $i_{d}=\hat{4}$, or $\hat{5}$;
$v_{c}=1$ when $i_{c}=\hat{1}$, $\hat{3}$, or $\hat{5}$; and
$v_{d}=1$ when $i_{d}=\hat{1}$, $\hat{3}$, or $\hat{5}$.
All the exponents are $0$, otherwise.

\end{document}